\documentstyle[eqsecnum,aps,amstex,epsfig,float,amssymb]{revtex}

\begin{document}

\newcommand{\e}{\operatorname{e}}

% Special commands for bibliography
\newcommand{\aeta}{{\em Astron. Astrophys.}}
\newcommand{\astrophysj}{{\em Astrophys. Jour.}}
\newcommand{\mnras}{{\em Mon. Not. R. Ast. Soc.}}
\newcommand{\nuclphysa}{{\em Nucl. Phys. A}}
\newcommand{\physrevd}{{\em Phys. Rev. D}}
\newcommand{\physrevl}{{\em Phys. Rev. Lett.}}
\newcommand{\apjlet}{{\em Astrophys. Jour. Lett.}}

\draft

\title{An elliptical tiling method to generate a 2-dimensional set of templates for gravitational wave search}

\author{Nicolas Arnaud, Matteo Barsuglia, Marie-Anne Bizouard, Violette Brisson,\\Fabien Cavalier,Michel Davier, Patrice Hello, Stephane Kreckelbergh and Edward K. Porter}

\address{Laboratoire de l'Acc\'el\'erateur Lin\'eaire, B.P. 34, B\^atiment 200,
Campus d'Orsay, 91898 Orsay Cedex (France)\protect\\}

\maketitle

%================================================================
\begin{abstract}
%================================================================

Searching for a signal depending on unknown parameters in a noisy background with matched filtering techniques always requires an analysis of the data with several templates in parallel in order to ensure a proper match between the filter and the real waveform. The key feature of such an implementation is the design of the filter bank which must be small to limit the computational cost while keeping the detection efficiency as high as possible. This paper presents a geometrical method which allows one to cover the corresponding physical parameter space by a set of ellipses, each of them being associated to a given template. After the description of the main characteristics of the algorithm, the method is applied in the field of gravitational wave (GW) data analysis, for the search of damped sine signals. Such waveforms are expected to be produced during the de-excitation phase of black holes -- the so-called 'ringdown' signals -- and are also encountered in some numerically computed supernova signals. First, the number of templates $\mathfrak{N}$ computed by the method is similar to its analytical estimation, despite the overlaps between neighbor templates and the border effects. Moreover, $\mathfrak{N}$ is small enough to test for the first time the performances of the set of templates for different choices of the minimal match $MM$, the parameter used to define the maximal allowed loss of signal-to-noise ratio (SNR) due to the mismatch between real signals and templates. The main result of this analysis is that the fraction of SNR recovered is in average much higher than $MM$, which dramatically decreases the mean percentage of false dismissals. Indeed, it goes well below its estimated value of $1-MM^3$ used as input of the algorithm. Thus, as this feature should be common to any tiling algorithm, it seems possible to reduce the constraint on the value of $MM$ -- and indeed the number of templates and the computing power -- without loosing as much events as expected in average. This should be of great interest for the inspiralling binaries case where the number of templates can reach some hundreds of thousands for the whole parameter space.

%================================================================
\end{abstract}
%================================================================


\baselineskip = 2\baselineskip 

%================================================================
\section{Introduction}
%================================================================

In the next years, the first generation of large interferometric gravitational-wave detectors \cite{ligo,virgo,tama,geo,aciga} should reach a sensitivity good enough to expect the first direct detection of GW signals. In parallel of the experimental work consisting in operating the detectors at their working point with background noises as small as possible, the future data analysis methods are being prepared for the various expected sources of GW, each of them requesting specific tools. Yet, the Wiener filtering is used in most of these fields due to its 'optimal' characteristics for signals whose time evolutions are known. Indeed, it is not only the filter giving the highest signal-to-noise ratio (SNR) among all the linear ones \cite{WZ} but it has also the property of having the lowest false dismissal rate for a given false alarm rate among all filters \cite{MF_optimal}.

Conversely, its main drawback is its poor robustness: as soon as the physical signal and the filter do not match exactly, the SNR can be dramatically reduced. Even if the searched waveform is analytically computed with high precision, it always depends on a vector of parameters\footnote{By parameters we mean here the {\it intrinsic} ones; we assume that all the {\it extrinsic} variables such as the signal timing arrival can be properly maximized numerically.} $\vec{\lambda}$ whose values are specific to a given source and thus unknown -- e.g. its mass, the main frequency of emission... -- and whose accurate estimation is indeed a major aim of the data analysis.

The set of physically possible values for the vector $\vec{\lambda}$ is the continuous parameter space $\mathbb{P}$. A given filter can only be 'efficient' 
-- i.e. recovering a large fraction of the SNR -- in a restricted region of this space, called the efficiency area. Therefore, many different templates must be used in parallel for a matched filtering procedure, in order to cover the whole parameter space. The lattice must be dense enough to ensure a minimal loss in SNR for any real signal whose parameters do not exactly match any of the available filters, while the number of templates $\mathfrak{N}$ must be kept small to limit the computing power needed for the analysis.

Given these two requirements, it clearly appears that the choice of the set of templates is very important as it has major consequences on both the ultimate filtering performances and its feasibility. On the other hand, 'tiling' $\mathbb{P}$ is most of the time very difficult as the efficiency area ${\mathbb{E}}$ of a filter depends on $\vec{\lambda}$ and on the minimal match $MM$ -- see Section \ref{subsection:formalism}. Thus, in the general case, uniform coverage is not possible. 

In the one-dimensional case ($D_{\mathbb{P}}=1$) the tiling is easy as the efficiency areas are straight segments of various lengths which can be aligned one after the other one -- see e.g. \cite{these_nico} where the matched filtering with Gaussian peak templates (only depending on their width) is studied. As soon as $D_{\mathbb{P}} \ge 2$, one has to cope with overlaps, holes and borders. 

The problem of computing a set of templates for matched filtering purpose has already been studied for the compact inspiralling binaries search. The GW signal is accurately estimated thanks to various development methods -- Taylor \cite{Taylor} or Pade \cite{Pade} post-Newtonian (PN) expansions -- and both the estimated SNR and the event rate make such events good candidates for a first detection. If the spins are neglected and the orbit circular, the parameter space is two-dimensional (the two masses of the stars) and the number of templates can be very high \cite{porter} if the low mass region is included in $\mathbb{P}$.

Using the formalism defined in Ref.\cite{Owen}, Owen and Sathyaprakash present a method \cite{Owen_Sathya} to cover this parameter space at the 2 PN order. To solve the question of placing new filters with respect to the previous ones, rectangles inscribed inside the efficiency areas are used instead of the real ellipses, bigger but more difficult to place accurately -- i.e. without holes. 

In this paper, a different tiling approach is studied for the case $D_{\mathbb{P}} = 2$. The main idea is to construct the lattice of filters with an iterative method ensuring a (rather optimal\footnote{In the sense that the local number of templates is minimized for circular and slowly varying efficiency areas.}) {\it local} template placement. The location of the first ellipse center is arbitrary and chosen by the user. Then the procedure goes on and new ellipses increasing the coverage of $\mathbb{P}$ are added one after the other. When it stops, the tiling of the full parameter space is normally complete -- $\mathbb{P}$ fully covered -- but is certainly not optimal from the {\it global} point of view. As described in Section \ref{section:tiling_details}, the overlapping in the final configuration can be so important that a large fraction of ellipses are in fact completely covered by other ones and thus useless for the detection purposes, while wasting computing time.

Therefore, a second step is then started to clean the list of templates by selecting these ellipses and erasing the corresponding filters. In the end, the final set of templates -- all useful -- only depends on the position of the first filter in the parameter space from which the locations of all the other ones have been iteratively computed. Due to border effects which cannot be properly estimated, it is impossible to define the best initial position. So, a last trick to further decrease the number of templates is to merge some sets of templates corresponding to different starting points and to apply to the full list the reduction procedure previously mentioned. 

As our team is mostly involved in the search for GW bursts (short GW signals usually lasting a few milliseconds for which waveforms cannot be predicted as accurately as for coalescing binaries, emitted by e.g. supernovae or the merging phase of compact stars inspiralling one around the other), this algorithm was originally developed to search for damped sine-like signals. Indeed, such time-behavior can be seen in a large fraction of supernova explosion GW signals computed numerically -- see e.g. the waveforms of Ref.\cite{SN_num} -- and is also expected to occur when excited black holes come back to equilibrium \cite{Press,Teukolsky_Press}. The latter case will be used in the following as a benchmark of the tiling method performances; details are presented in Section \ref{section:damped_sine}.

Another interesting feature of this example is that the number of templates remains small even when the allowed mismatch loss of SNR $1-MM$ is kept very low. Thus, it is possible to check the quality of the tiling (with respect to the prescription originating the tiling method) and to see how the characteristics of the set of templates evolve when $MM$ changes. This experience may be very useful for the inspiralling binaries search whose tilings cannot be tested so easily -- yet, they must show similar behaviors.

In particular, the most significant result we obtain is that one should be able to reduce the constraint on the template spacing while keeping small the mean false dismissal rate of events, ultimately the only important quantity in the search of rare signals occurring at random time. Therefore, a much less numerous set of filters would still be efficient enough, but at a smaller computing cost.

%============================================================================
\section{Details on the tiling algorithm}
%============================================================================
\label{section:tiling_details}

After the introduction of some hypothesis and useful notations, the main steps of the tiling method are presented.

%================================================================
\subsection{Tiling formalism}
%================================================================
\label{subsection:formalism}

%================================================================
\subsubsection{Hypothesis and notations}
%================================================================

As a linear filtering consists in correlating the detector output $s(t)$ with the corresponding filtering function $\varphi(t)$, one defines the following scalar product to represent the filtering operation:

\begin{equation}
\langle \, s \, | \, \varphi \, \rangle \; = \; 4 \; \Re \left( \int_{0}^{\infty} \, df \, 
\frac{ \tilde{s}(f) \, \tilde{\varphi}(f)^{*} }{ S_h(f) } \right)
\label{eq:inner_product_frequency}
\end{equation}

where $S_h$ is the one-sided power spectrum density and the ``$\tilde{~}$'' symbol means Fourier transform. The normalization is chosen so that, in case of signal alone, $\sqrt{\langle s | s \rangle}$ is equal to the SNR.

If one now assumes that $\varphi(t)$ is almost monochromatic signal of frequency $f_0$, one can show that if the noise spectrum is nearly flat around $f_0$, Eq. (\ref{eq:inner_product_frequency}) becomes approximately equal to the following equation:

\begin{equation}
\langle \, s \, | \, \varphi \, \rangle \; = \; \frac{2}{ S_h(f_0) } \; \int_{\mathbb{R}}
 \, s(t) \, \varphi(t) \, dt
\label{eq:inner_product}
\end{equation}

As the definition of the ambiguity function involves normalized templates -- see Section \ref{subsubsection:ambiguity} --, the frequency-depending factor $S_h(f_0)$ vanishes. Thus, representing the filtering operation by a scalar product in the time domain is in fact accurate also with a colored noise for signals with a narrow extension in the frequency domain. This will be approximately the case for the damped sine waveforms -- see Ref.\cite{Flanagan_Hughes} for a detailed discussion -- which will be used to test the tiling algorithm in the next sections.

%================================================================
\subsubsection{Ambiguity function}
%================================================================
\label{subsubsection:ambiguity}

The ambiguity function $\Gamma$ between two templates $k_{\vec{\lambda}}$ et $k_{\vec{\lambda} + d\vec{\lambda}}$ is defined by:

\begin{equation}
\Gamma\left( \vec{\lambda} \; ; \; d\vec{\lambda} \right) = \langle \; k_{\vec{\lambda}} \; | \; k_{\vec{\lambda} + d\vec{\lambda}} \; \rangle
\end{equation}

The templates are normalized: $\langle \, k_{\vec{\lambda}} \, | \, k_{\vec{\lambda}} \, \rangle = 1 = \langle \, k_{\vec{\lambda} + d\vec{\lambda}} \, | \, k_{\vec{\lambda} + d\vec{\lambda}} \, \rangle$. $\Gamma$ is thus a measurement of the closeness between two templates. It can also considered as a way to see how well the template $k_{\vec{\lambda}}$ can be used to detect the signal $k_{\vec{\lambda} + d\vec{\lambda}}$: the ambiguity function is the mean fraction of the optimal SNR achieved at a given distance in parameter space.

Following Ref.\cite{Owen}, if $d\vec{\lambda}$ is small enough, the ambiguity function can be approximated by a second order power expansion -- the first order is null as the expansion is performed around the absolute maximum of $\Gamma$.

\begin{equation}
%\Gamma\left( \vec{\lambda} \; ; \; d\vec{\lambda} \right) \; = \; 1 \; - \; \langle \; d\vec{\lambda} \; | \; {\mathfrak{M}}(\vec{\lambda}) \; | \; d\vec{\lambda} \; \rangle
\Gamma\left( \vec{\lambda} \; ; \; d\vec{\lambda} \right) \; = \; 1 \; - \; \frac{1}{2} \, g_{\mu\nu} \, d\lambda^\mu \, d\lambda^\nu
\label{eq:metric}
\end{equation}
%where ${\mathfrak{M}}(\vec{\lambda})$ is the matrix of a positive definite quadratic form.
with $g_{\mu\nu}$ defining a metric on the parameter space $\mathbb{P}$ \cite{bala,Owen}.

Finally, one defines the minimal match $MM$ as the lower bound of the recovered SNRs, which means that the loss of the SNR due to the mismatch between the template and the signal must be kept below $1-MM$. Following Ref.\cite{Owen}, one can note that with this definition, a -- of course {\it pessimistic} -- estimator of the fraction of false dismissals $\mathfrak{L}$ is:

\begin{equation}
{\mathfrak{L}} \; = \; 1 \, - \, MM^3
\end{equation}

For instance, with $MM=97\%$, one has ${\mathfrak{L}} \sim 10\%$. This value of $MM$ is usually found in the literature as the correspondence with this particular value of $\mathfrak{L}$ is easy.

This quantity $MM$ is the only input parameter of the tiling procedure; it allows one to define precisely the efficiency area ${\mathbb{E}}(\vec{\lambda},MM)$ of the template $k_{\vec{\lambda}}$ by the following equation:

\begin{equation}
%\langle \; d\vec{\lambda} \; | \; {\mathfrak{M}}(\vec{\lambda}) \; | \; d\vec{\lambda} \; \rangle \; \le \; 1 \, -\, MM
\frac{1}{2} \, g_{\mu\nu} \, d\lambda^\mu \, d\lambda^\nu \; \le \; 1 \, -\, MM
\end{equation}
The area of $\mathbb{P}$ including all the vectors $k_{\vec{\lambda} + d\vec{\lambda}}$ which match this inequality is the inner part of an ellipsoid centered on $\vec{\lambda}$ whose proper volume scales as $( 1 - MM )^{-D_{\mathbb{P}}/2}$. The average fraction of recovered SNR for physical signals with parameters belonging to ${\mathbb{E}}(\vec{\lambda},MM)$ is at least equal to $MM$.

%============================================================================
\subsubsection{The tiling problem}
%============================================================================

Tiling the parameter space consists of finding a set of ${\mathfrak{N}}$ filters 
$\left( k_{\vec{\lambda_p}}\right)_{1 \le p \le {\mathfrak{N}}}$ so that the union of the surfaces $\left({\mathbb{E}}(\vec{\lambda_p},MM)\right)_{1 \le p \le {\mathfrak{N}}}$ completely covers $\mathbb{P}$. The aim is to achieve this task by using as few templates as possible to keep the computing cost manageable. To do so, one has to solve two related questions: 
\begin{itemize}
\item which tiling algorithm to use?
\item how to test its quality?
\end{itemize} 

The main problems encountered by any tiling procedure are: overlapping, gaps, areas lost beyond the physical parameter space... The difference between the ellipsoid and the real border of the efficiency area is not really taken into account: in practical cases for one-step searches, $MM$ is close to 1 and thus the ellipsoid is assumed to be a good approximation of the ambiguity surface. Moreover, the overlapping between close ellipses is an advantage to discard such a problem. In fact, the tests of the sets of templates constructed with the tiling  method presented below show that the tilings have no significant holes, even with the choice $MM=0.85$, which validates a posteriori this hypothesis. Clearly, the value of $MM$ from which the ellipsoid approximation of the efficiency area is no longer valid depends on the precise tiling problem considered.

%============================================================================
\subsection{Covering the space parameter}
%============================================================================

It is well-known that the optimal tiling of an infinite plane by identical disks of radius $R$ consists in placing their centers on an hexagonal lattice, separated by a distance of $\sqrt{3} R$ -- see Figure \ref{fig:optimal_circle_tiling}. In this way, the overlapping is minimized and the ratio $\eta_{\text{opt}}$ between the sum of the disk areas and the surface effectively covered is

\begin{equation}
\eta_{\text{opt}} \; = \; \frac{ 2 \, \pi }{ 3 \, \sqrt{3} } \; \approx \; 1.21
\end{equation}
This value would increase in any real situation due to the border effect.

This property of circular tiling is used by the algorithm on the following way. Once the location of a template $\vec{k}_0$ in $\mathbb{P}$ has been chosen, one computes its efficiency area ${\mathbb{E}}_0$. Through a simple plane transformation, ${\mathbb{E}}_0$ becomes a circle ${\mathbb{C}}_0$ of unit 
radius  and the six neighbor centers are placed on the regular hexagonal lattice previously described. This choice is nearly optimal if the shape of the efficiency areas is slowly varying with respect to their characteristic dimensions. Using the inverse transformation, the six new center positions are given in $\mathbb{P}$. Then, each ellipse associated with a new center is tested in order to check if it covers a part of $\mathbb{P}$ not already covered by the previously defined ellipses. If the ellipse is useful, its center is added to the center list and will be used later to place other centers. For example, we have found that with $MM=97\%$ only about 20\% of the centers are kept. This iterative algorithm stops either when the full surface of the parameter space is covered or when no new ellipse can be placed any longer.

%============================================================================
\subsection{Cleaning the template list}
%============================================================================

A successfully completed coverage computed with the former procedure can be usually redundant in the sense that a large fraction of ellipses are completely covered by others. To save computing time, useless templates need to be identified and deleted. Like for the first step of the algorithm, one has to face the problem that erasing an ellipse is a local operation which can have global consequences: which template should be discarded first?

To answer this question, the following procedure has been set:

\begin{itemize}
\item For each ellipse $\mathbb{E}$ belonging to the tiling, one defines its {\it utility} $\Upsilon({\mathbb{E}})$ which is the fraction of $\mathbb{P}$ it covers alone. Useless ellipses are thus characterized by $\Upsilon = 0$.
\item Among those ellipses, the one with the smallest area in the parameter space is dropped. This prescription is consistent with the idea that it is a priori better to keep big ellipses which have a  higher potential of coverage. Of course this general rule may not always be true but it looks reasonable.
\item Utilities are then updated for the surrounding ellipses as they become more ``useful'' -- $\Upsilon({\mathbb{E}})$ locally increases. Then, this scheme is iterated until all the remaining ellipses have non zero utilities.
\end{itemize}

This second step of the tiling algorithm is very important: for most of the simulations in the damped sine case, the number of templates is reduced by a factor of about two with respect to the first list.

%============================================================================
\subsection{Merging template lists and final cleaning}
%============================================================================
\label{subsection:merging}

Finally, the output of the whole procedure is a set of templates completely covering the parameter space and which depends on only one initial condition, that is the location of the first filter set at the beginning of the tiling generation. Once more, this choice may have global consequences on the tiling quality -- for instance, shifting it 'horizontally'/'vertically' may cause  a column/line of templates to appear or vanish -- but cannot be simply optimized. Therefore, tilings starting at different points on the parameter space are computed in parallel; then, all the lists of templates are merged in a single one. Finally, the cleaning procedure is applied to this clearly redundant set of filters to obtain the final lattice of templates. Thanks to this last step, its size is decreased by 10 or 15\% with respect to the individual coverages in the damped sine case. 

In the damped sine case, five different tilings have been merged together to compute the final lattice of templates. There seems to be a good compromise between gain in the number of templates and computation time (more than 80\% of the ellipses of the full list are useless and so the cleaning procedure is much longer). This last merging step doubles the total computation time of the tiling, i.e. its duration is of the same order of magnitude as the sum of the computation times for the initial tilings.

%============================================================================
\subsection{Estimation of the tiling quality}
%============================================================================

Three variables can be used to control the quality of the tiling.

\begin{itemize}
\item The number of templates $\mathfrak{N}$.
\item The ratio $\eta_{\text{tot}}$ between the sum of all the ellipse areas and the parameter space surface.
\item The ratio $\eta_{\text{in}}$ between the sum of all the ellipse areas inside $\mathbb{P}$ and the parameter space surface.
\end{itemize}

One has clearly $\eta_{\text{in}} \le \eta_{\text{tot}}$. These two estimators allow to measure the overlapping between templates and the fraction of extra area outside the parameter space, while $\mathfrak{N}$ can be compared with its estimation computed by integrating the proper volume of the parameter space. Due to the imperfections of the real tiling (overlapping, border effects...) and to the fact that the analytical computation of the template numbers is only approximative by principle, it is interesting to check the consistency of the two numbers.

%================================================================
\section{Matched filtering detection of damped sinusoidal signals}
%================================================================
\label{section:damped_sine}

%================================================================
\subsection{Normal modes of black hole oscillations}
%================================================================

An excited black hole, born e.g. after a supernova collapse or the merging of two compact objects, comes back to a stationary state by emitting GW. This emission can be described as a superposition of black hole quasi-normal modes \cite{Press,Teukolsky_Press}. The dominant mode is expected to be quadrupolar with the longest damping time \cite{Echevarria}. In order to limit the number of free parameters defining the matched filtering parameter space, one assumes that after some transitory phase the waveform becomes (with a proper choice of time origin):

\begin{equation}
h(t) \; \propto \; \exp\left(-\frac{t}{\tau}\right) \; \sin( 2 \pi f t) 
\end{equation}

It was observed \cite{Detweiler} that a given couple $(f,\tau)$ is connected to one single set of physical parameters $\left(M_{\text{BH}},a_{\text{BH}}\right)$, the mass and the reduced angular momentum of the black hole. Moreover, the corresponding relation can be expressed analytically with a 10\% precision \cite{Echevarria,Leaver}. Introducing the quality factor $Q = \pi f \tau$, one has in geometrical units ($G = c = 1$):

\begin{eqnarray}
\label{eq:Q}
Q \; &\approx& \; 2 \, ( 1 \, - \, a_{\text{BH}} )^{-9/20} \\
\label{eq:f}
f \; &\approx& \; \frac{1}{2 \, \pi \, M_{\text{BH}}} \; \left[ 1 \, - \, 0.63 \, \left( 1 \, - \, a_{\text{BH}} \right)^{3/10}\right]
\end{eqnarray}

The normal mode frequency is, as expected, a decreasing function of the black hole mass, increasing with the rotation parameter as the quality factor, which is also independent of $M_{\text{BH}}$. Therefore, detecting such GW signal would give direct physical information on its source.

%================================================================
\subsection{Signal to noise ratio}
%================================================================

Following Ref.\cite{Flanagan_Hughes}, two different methods estimating the optimal SNR give identical results:

\begin{itemize}
\item a calculation in the direct space assuming that the noise is white at the oscillation frequency;
\item computing the Fourier spectrum and approximating it by a Dirac at the oscillation frequency.
\end{itemize}
One has finally

\begin{equation}
\rho_{\text{max}} \; \propto \; \frac{ M_{\text{BH}} }{ \text{distance} } \; \sqrt{ \frac{ Q }{ f \, \times \, S_h( f ) } } 
\end{equation}
where $S_h$ is the one-sided power spectrum density of the noise. The proportionality constant depends on the physical process at the origin of the black hole formation. For high mass inspiralling binaries, it can be very large \cite{Flanagan_Hughes} and leads to detections at cosmological distances, while for supernova collapses, it can be high enough to be detected in the Local Group \cite{SP,Ferrari_Palomba}.

Using this framework, it is sufficient to study the two quadratures and to compute the corresponding sets of templates covering the two dimensional parameter space ${\mathbb{P}} = \{ Q, f \}$.

%================================================================
\subsection{Tiling the two dimensional parameter space}
%================================================================

As in the black hole oscillation case, no correlation is assumed between the oscillation frequency $f$ and the quality factor $Q$. Therefore, $\mathbb{P}$ is rectangular: 

\begin{equation}
\nonumber
{\mathbb{P}} \; = \; \left[ Q_{\text{min}} \, ; \, Q_{\text{max}} \right] \; \times \; \left[ f_{\text{min}} \, ; \, f_{\text{max}} \right ]
\end{equation} 

Its borders are chosen in the following way:

\begin{itemize}
\item For Q, by using the black hole normal mode range \\
\begin{equation}
\nonumber
\begin{cases}
Q_{\text{min}} \; = \; Q( a_{\text{BH}} = 0 ) = 2 \\
Q_{\text{max}} \; = \; Q( a_{\text{BH}} = 0.99 ) \approx 16
\end{cases}
\end{equation}
\item For f, by using the interferometric detector main characteristics \\
\begin{equation}
\nonumber
\begin{cases}
f_{\text{min}} \; = \; 20 \; \text{Hz} \;\;\; \text{(roughly twice the realistic value for the Virgo seismic wall)} \\
f_{\text{max}} \; = \; 10 \; \text{kHz} \;\;\; \text{(the Nyquist frequency corresponding to the Virgo sampling frequency of $f_{\text{samp}} = 20$ kHz)} 
\end{cases}
\end{equation}
\end{itemize} 

The mismatch between two templates of parameters $(Q_0,f_0)$ and $(Q=Q_0+\delta Q, f=f_0 + \delta f$) can be written\footnote{From the usual definition of the metric on the parameter space -- see Eq. (\ref{eq:metric}) -- there is a factor 2 between the ellipse coefficients and the metric ones: $g_{QQ} = 2 \alpha/ Q^2$, $g_{Qf} = 2 \beta / Qf$ and $g_{ff} = 2 \gamma / f^2$.}

\begin{equation}
\alpha \ \left( Q_0, f_0 \right) \, \left( \frac{ \delta Q }{ Q_0 } \right)^2 \; + \; 2 \, \beta \, \left( Q_0, f_0 \right) \, \left( \frac{ \delta Q }{ Q_0 } \right) \, \left( \frac{ \delta f }{ f_0 } \right) \; +\; \gamma \, \left( Q_0, f_0 \right) \, \left( \frac{ \delta f }{ f_0 } \right)^2 \; \le \; 1 \, - \, MM
\end{equation}
with $1-MM$ being the maximal expected loss in SNR. The dimensionless coefficients $\alpha$, 
$\beta$ and $\gamma$ only depend on the quality factor $Q$. Their expressions are the following:

\begin{itemize}
\item for pure damped cosines:
\begin{eqnarray}
\nonumber
\alpha( Q ) &=& \frac{1}{8} \; \frac{ 64Q^8+128Q^6+28Q^4+1 }{ (1+4Q^2)^2 \, (1+2Q^2)^2 } \\
\nonumber
\beta( Q ) &=& -\frac{1}{8} \; \frac{ 8Q^4+2Q^2+1 }{ (1+2Q^2) \, (1+4Q^2) }\\
\nonumber
\gamma( Q ) &=& \frac{1}{8} \; \frac{ 16Q^4+6Q^2+1 }{ 1+2Q^2 }
\end{eqnarray}
\item for pure damped sines:
\begin{eqnarray}
\nonumber
\alpha( Q ) &=& \frac{1}{8} \; \frac{ 16Q^4+3 }{ (1+4Q^2)^2 } \\
\nonumber
\beta( Q ) &=& -\frac{1}{8} \; \frac{ 3+4Q^2 }{ 1+4Q^2 }\\
\nonumber
\gamma( Q ) &=& \frac{8Q^2+3}{8}
\end{eqnarray}
\end{itemize}

In the former case, assuming that the quality factor is much greater than 1, the first order expansion of these coefficients in power of $Q$ gives the results computed in Ref.\cite{Creighton}. The hypothesis $Q \gg 1$ is not valid in the total range $\left[Q_{\text{min}} = 2 \, ; \, Q_{\text{max}} = 16 \right]$ we have considered in this paper and the difference between the real coefficients and their first order approximations can reach 10\% in the region of low quality factor.

The number of templates needed to cover the parameter space can be roughly estimated by the 
usual formula:
\begin{equation}
{\mathfrak{N}}  \, \sim \,  \frac{ 1}{ \mathcal{V}} \int_{ {\mathbb{P}}} 
\, \sqrt{g} \, dx^{\mu} dx^{\nu} = \,  \frac{ 2}{ \mathcal{V}} \int_{ {\mathbb{P}}} 
\,  \sqrt{ \alpha\gamma \, - \, \beta^2} \, \frac{ dQ_0 }{ Q_0 } \, \frac{ df_0 }{ f_0 },
\end{equation}
where ${ \mathcal{V}}$ is the template proper volume and $g$ is the metric determinant. 
As ${ \mathcal{V}} \sim 4 \, ( 1 - MM )$, we can express this number
with the help of $MM$:
\begin{equation} 
{\mathfrak{N}}  \sim  \frac{ 1 }{2 \, ( 1 - MM ) } \, \int_{ {\mathbb{P}} } 
\, \sqrt{ \alpha\gamma \, - \, \beta^2} \, \frac{ dQ_0 }{ Q_0 } \, \frac{ df_0 }{ f_0 }.
\label{eq:nb_templates}
\end{equation}

From Equation (\ref{eq:nb_templates}), it is possible to 
estimate the numbers of templates $\mathfrak{N}$ needed to cover the parameter space. One gets
at the first order in Q: 
\begin{equation}
{\mathfrak{N}} \, \sim \, \frac{ \sqrt{ 2 } }{ 8 \, ( 1 \, - MM ) } \, 
\ln\left( \frac{ f_{\text{max}} }{ f_{\text{min}} } \right) \, \left( Q_{\text{max}} 
\, - \, Q_{\text{min}} \right)
\label{eq:N_owen}
\end{equation}
for the two families of signals studied. The number of templates is logarithmic in the frequency band and scales linearly in the quality factor range. With the numerical border values previously defined, one gets ${\mathfrak{N}} \approx 500$.

The expressions giving the coefficients $\alpha$, $\beta$ and $\gamma$ clearly show a dependence on
 the ellipse location in the parameter space. Using the vocabulary of differential geometry, this is a strong indication that $\mathbb{P}$ is curved. Following Ref.\cite{Ed_Stas}, it is possible to verify this by calculating the curvature of the search manifold.  As the waveform is a function of the parameters $\{Q,f_{0}\}$ only, we have a two dimensional problem and can thus calculate the Gaussian curvature, $K$, directly.  Following the Petrov classification, the Riemann tensor $R^{\alpha}_{\,\,\,\beta\mu\nu}$ associated with the metric defined in Eq. (\ref{eq:metric}) has only one independent component, which we can take to be $R_{1212}$, in its fully covariant form.  We can thus define the Gaussian curvature on the search manifold by
\begin{equation}
K \; = \; \frac{ R_{1212} }{ g },
\end{equation}
where $g$ is the metric determinant. For the damped sine case for instance, the Gaussian curvature is constant over $P$: $K = 4$.  Therefore, the search manifold has the same topology as a 2-sphere.

%================================================================
\subsection{Template number}
%================================================================

As the template tilings appear to be very similar for the damped cosine and sine signals -- compare for instance the expressions giving the coefficients $\alpha$, $\beta$ and $\gamma$ --, we mainly focus on the latter in the following.

\begin{center}
\begin{tabular}{|c|c|c|c|c|}
\hline Starting point ($Q$, $f$) & Template number before cleaning procedure & Final template number $\mathfrak{N}$ & $\eta_{\text{in}}$ & $\eta_{\text{tot}}$ \\
\hline (2, 20 Hz) & 1992 & 854 & 2.05 & 2.57 \\
\hline (9, 4990 Hz) & 1570 & 873 & 2.08 & 2.61 \\
\hline (16, 10 kHz) & 1532 & 888 & 2.11 & 2.68 \\
\hline (14, 100 Hz) & 1986 & 849 & 2.04 & 2.51 \\
\hline (3, 9900 Hz) & 1491 & 857 & 2.15 & 2.74 \\
\hline
\end{tabular}
\end{center}
\centerline{Table VII: Characteristics of tilings computed from different starting points}

\begin{figure}
\centerline{\epsfig{file=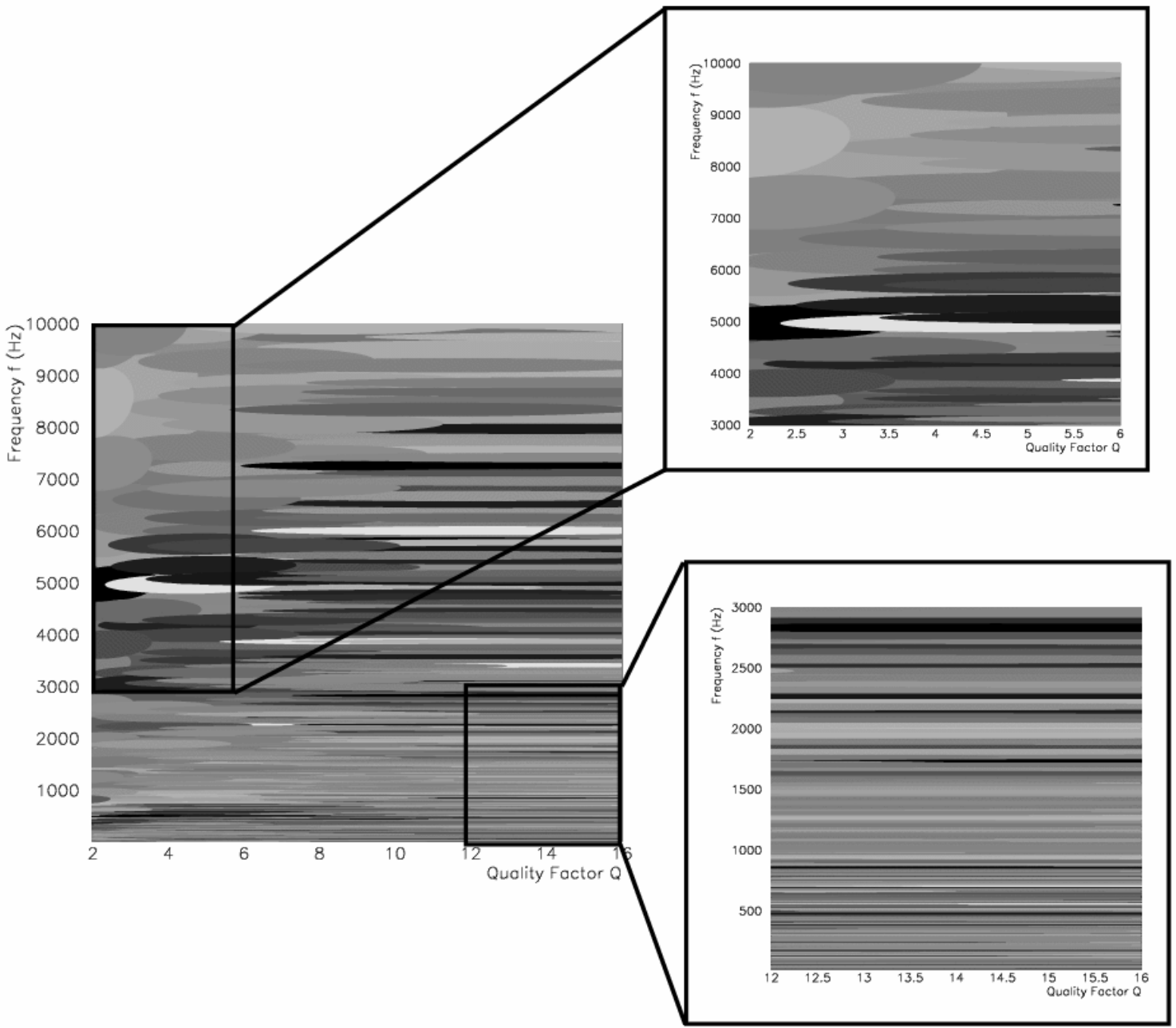,width=20cm}}
\caption{Zooms of the tiling in two different regions of the parameter space: one can clearly see the large variations in the ellipse shape in $\mathbb{P}$.}
\label{fig:tiling_zoom}
\end{figure}

Table VII shows the main characteristics of some parameter space coverages computed with the tiling procedure presented in this paper, starting from different initial points in $\mathbb{P}$. The number of templates generated by the first phase of the algorithm to cover a given parameter space depends greatly on the initial point chosen in $\mathbb{P}$: one can see variations of about 20\%. By comparing the number of templates before and after reduction, the importance of the cleaning procedure becomes clear: about 50\% of filters belonging to the initial set are finally rejected. The reduction by a factor 2 was surprising even though it was supposed that each ellipse is placed in a nearly optimal way. In fact, for the damped sine case, the assumption that the shape of the efficiency areas are slowly varying with respect to their characteristic dimensions is far from true as shown in the zooms of Figure \ref{fig:tiling_zoom}. %In order to quantify the shape variations, the evolution on $\mathbb{P}$ of the scalar curvature is plotted on Figure ??? (nouvelle figure). 

%??? Commentaire a faire en fonction de la figure ???

Border effects play a limited -- although non zero -- role for the size of the tiling as $\mathfrak{N}$ changes by a few percent (between 5 and 10\% at most) with the position of the initial templates. Yet, comparing $\eta_{\text{in}}$ and $\eta_{\text{tot}}$ allows one to see that a large fraction of the ellipse potential -- i.e. not taking into account overlaps -- areas is lost outside of $\mathbb{P}$. Moreover, the values of the two 'quality' estimators $\eta_{\text{in}}$ and $\eta_{\text{tot}}$ remain much higher than for the optimal disk tiling which shows that ellipses do not tile in the best way. 

With these different tilings, the merging procedure presented in Section \ref{subsection:merging} has been applied. As expected it allows one to decrease $\mathfrak{N}$ by about 15-20\% more -- see Table VIII. The reason for this gain can be understood by looking at the two estimators $\eta_{\text{in}}$ and $\eta_{\text{tot}}$: their values decreased in the final step. Therefore, the template efficiency areas have less overlaps and better match the border of the parameter space.

\begin{center}
\begin{tabular}{|c|c|c|c|c|}
\hline & Full list size & $\mathfrak{N}$ after final cleaning & $\eta_{\text{in}}$ & $\eta_{\text{tot}}$ \\
\hline Damped sines & 4341 & 698 & 1.74 & 2.12 \\
\hline Damped cosines & 4346 & 720 & 1.87 & 2.26 \\
\hline
\end{tabular}
\end{center}
\centerline{Table VIII: Characteristics of the best results achieved by merging together lists of templates with different starting points.}

The last point to be mentioned here is that even if the ellipse characteristics change a lot in the parameter space, the algorithm succeeds in tiling the whole $\mathbb{P}$. Figure \ref{fig:tiling_zoom} presents two zooms of the coverage; the first one (top) corresponds to a region with small quality factors $Q$ and quite high frequencies $f$ where ellipses are large. The second (bottom) focuses on the area where $Q$ is high and $f$ below 3 kHz; there, the ellipses are very narrow in the frequency direction, which is simply due to the fact that a sine-like signal of 'small $f$' and high quality factor must be tracked with a very good accuracy: the frequency difference between the template and the signal must remain very low. Ultimately, at $Q \rightarrow +\infty$, the correlation would be a delta-function in the frequency difference. 

%================================================================
\subsection{Tests on the sets of templates}
%================================================================

A way to estimate the quality of a set of templates computed by the tiling algorithm is simply to check how it fulfills the input requirement of minimizing the loss of SNR for the detection of any signal belonging to the parameter space. Therefore, assuming a uniform distribution of the signal parameters in $\mathbb{P}$, Monte-Carlo simulations have been performed in order to compute the distribution of the fraction of SNR recovered by matched filtering with the two banks of filters. For each signal with random parameters, correlations are computed with all the templates and the higher match is selected. Ambiguity functions are first maximized on the time shift between the signal and the filter -- both sampled at the Virgo sampling frequency, $f_{\text{samp}} = 20$ kHz. No additional noise is needed for this optimization as one wants to estimate the mean loss of SNR.

\begin{figure}
\centerline{\epsfig{file=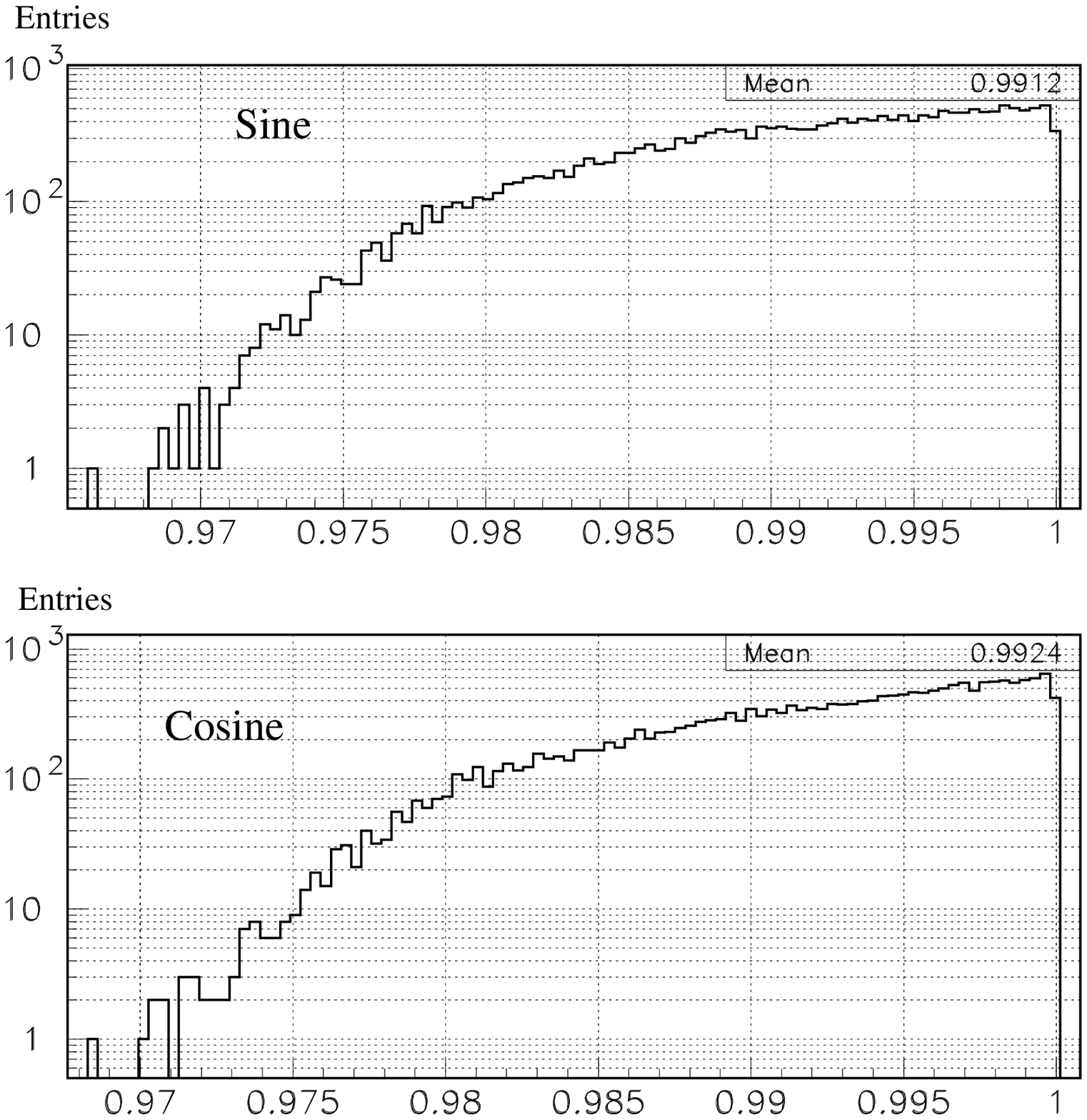,width=20cm}}
\caption{Distribution of the fraction of SNR recovered by the bank of templates for physical signals with unknown parameters, uniformly drawn in $\mathbb{P}$ ($MM=0.97$).}
\label{fig:ambiguity_distribution}
\end{figure}

The results can be seen on Figure \ref{fig:ambiguity_distribution}. By comparing them to the original value of the minimal match used to generate the sets of filters, $MM = 97 \%$, two important remarks can be made. First, the mean recovery fraction $\phi$ is 'much' higher than $MM$, more than 99\% for both the sine bank of templates and the cosine one. Then, a very small fraction of signals (less than 0.05\% in both cases) shows SNR losses higher than $1-MM$; yet, the recovered SNR does not go below 96.6\%, a value very close to $MM$. 

So, the overlapping between templates seen both on Figure \ref{fig:tiling_zoom} and through the high values of the estimators $\eta_{\text{in}}$ and $\eta_{\text{tot}}$ allows a better match between filters and signals. Moreover, the tilings show almost no 'holes' and guarantee an SNR loss lower than 3.4\%. Therefore, they appear satisfactory for data analysis purposes.

%================================================================
\subsection{Decreasing the minimal match $MM$}
%================================================================
\label{subsection:decreasing_MM}

If necessary, a way to decrease the number of templates would be to cover the parameter space with a less stringent requirement on the minimal match as the estimated number of templates scales as $(1-MM)^{-1}$ -- see Eq. \ref{eq:N_owen}. Indeed, the tilings presented in the previous sections with $MM=0.97$ show that in average the SNR loss remains much smaller than 3\%. Therefore, the mean estimated fraction of events lost because of mismatches between the finite set of templates and the physical GW signals $\overline{\mathfrak{L}}$ is well below ${\mathfrak{L}} = 1 - MM^3 \sim 10\%$. One has for the physical signals:

\begin{equation}
\overline{{\mathfrak{L}}} \; = \; 1 \; - \; \overline{ \phi^3_{\text{max} }}
\end{equation}
where the over-line represents an average over the GW signals and the maximum of $\phi$ is taken over all templates. One gets $\overline{{\mathfrak{L}}} \sim 2.3\%$ for the cosine filters and 2.6\% for the sine filters. Then tilings with smaller values of $MM$ are computed and their performances for the SNR recovery.

\begin{center}
\begin{tabular}{|c|c|c|c|c|c|c|c|}
\hline $MM$ & 97\% & 96\% & 95\% & 94\% & 92\% & 90\% & 85\% \\
\hline $\mathfrak{N}$ & 698 & 544 & 448 & 383 & 300 & 230 & 170 \\
\hline $\overline{\mathfrak{L}}$ (\%) & 2.6 & 3.3 & 4.2 & 4.8 & 6.2 & 7.5 & 10.7 \\
\hline
\end{tabular}
\end{center}
\centerline{Table IX: Template numbers and fraction of events lost for tilings with different values of the minimal match parameter $MM$}

\begin{figure}
\centerline{\epsfig{file=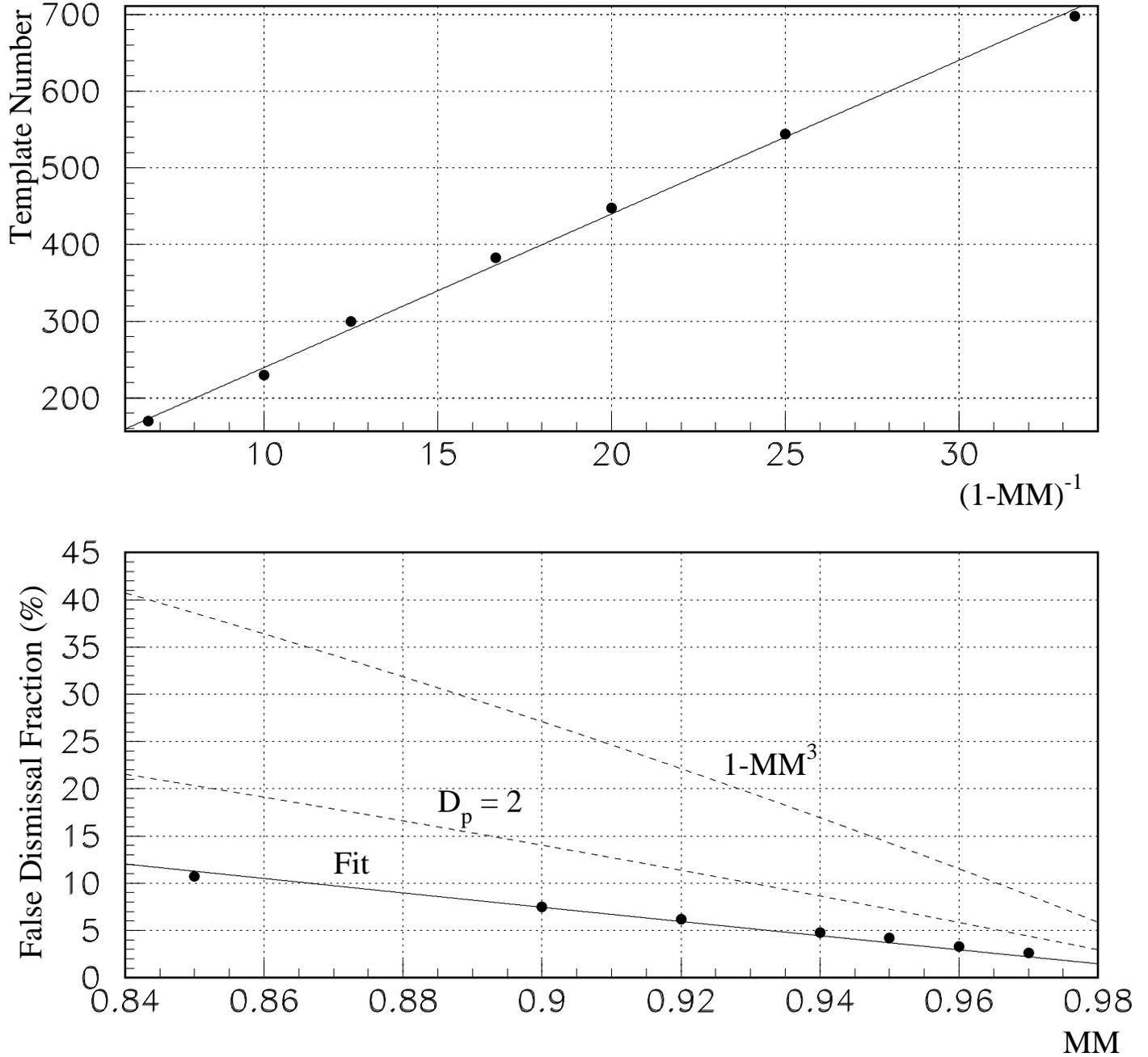,width=20cm}}
\caption{Top: evolution of $\mathfrak{N}$ versus $1/(1-MM)$ showing the expected linear behavior. Bottom: false dismissal fraction versus the minimal match $MM$. Even for $MM=90\%$, the lost remains below the 10\% level -- assumed to be acceptable and at the origin of the common choice of $MM=0.97$. The dashed lines present two possible estimations of ${\mathfrak{L}}$: ${\mathfrak{L}}=1-MM^3$ and the parameterization computed with the model $D_{\mathbb{P}}=2$. Both increase much faster than $\overline{\mathfrak{L}}$ with $1-MM$; yet, the second one overestimates less the real data. In the two graphs, the continuous curves show the best fit of the data.}
\label{fig:tiling_characteristics}
\end{figure}

Table IX shows the results achieved for the case of sine filters with $MM$ values between 85\% and 97\%. The evolution of the tiling characteristics with respect to $MM$ can also be seen on Figure \ref{fig:tiling_characteristics}. For each value of the minimal match, a procedure similar to the $MM=97\%$ case has been followed: computation of 5 different tilings corresponding to different starting points and then merging of the template lists to reduce the number of filters needed.

The first thing to note is that $\mathfrak{N}$ clearly decreases with $MM$ and follows quite perfectly the expected behavior -- see Eq. (\ref{eq:nb_templates}). With respect to the 'theoretical' number of templates, $\mathfrak{N}$ is slightly larger, due to border effects. Yet, they affect in the same 'relative' way the tilings corresponding to different values of the minimal match and so the relation ${\mathfrak{N}} \propto (1-MM)^{-1}$ is preserved. The top graph of Figure \ref{fig:tiling_characteristics} shows this relation; the line corresponds to the best linear fit computed with the available data. One gets:
\begin{equation}
{\mathfrak{N}} \; \approx \; \frac{ 20 }{ 1 \, - \, MM } \; + \; 43
\end{equation}

The slope of the function is of the same order of magnitude as the estimated value given by Eq.\ref{eq:N_owen}: $\frac{ \sqrt{ 2 } }{ 8 } \ln\left( \frac{ f_{\text{max}} }{ f_{\text{min}} } \right) \left( Q_{\text{max}} - Q_{\text{min}} \right) \approx 15.4$.

The second result -- and perhaps the most important -- is that even with $MM=0.90$ the estimated fraction of lost events is smaller than the expected 10\%, as shown on the bottom graph of Figure \ref{fig:tiling_characteristics}. With a template number reduced by a factor of 3, the parameter space coverage remains well inside the initial specification: $\overline{\mathfrak{L}} \le 10\%$. It is only with the choice $MM=0.85$ that $\overline{\mathfrak{L}}$ slightly exceeds 10\%, but with a reduction of a factor 4 in the number of templates. Numerically, one obtains

\begin{equation}
\overline{\mathfrak{L}} \; \approx \; 1 \, - \, MM^{\kappa} \;\;\; \text{with} \;\;\; \kappa \, = \, 0.74
\end{equation}

This feature is of small importance here as the total number of templates $\mathfrak{N}$ is only a few hundred but would be essential for the compact inspiralling binaries searches. In this case, for a total mass of two solar masses, the number of templates is expected to be close to $10^5$ or even more depending on detector bandwidth -- see e.g. Ref.\cite{porter}. Any tiling method based on the minimal match criterion like the one developed in this paper should provide a set of templates able to recover on average a fraction of SNR much better than $MM$ -- the {\it Minimal} Match indeed --, and thus miss less events than assumed during the building phase of the lattice.

Therefore, by using a Monte-Carlo study of the generated tiling performances, it should be possible to decrease by a large fraction the number of templates really needed to guarantee that for instance 90\% of the events potentially detectable will be seen. Such a refinement appears to be a way to save a large amount of computing time and would simplify any data analysis strategy based on matched filtering methods.

%================================================================
\subsection{Discussion}
%================================================================

As known since many years, mean losses in SNR recovery are well below $1-MM$, which indeed implies that the mean fraction of lost events is much smaller than $1-MM^3$. In the previous section, the evolution of $1-\overline{\mathfrak{L}}$ was shown to be also well-fitted by a power law $MM^\kappa$, with the exponent $\kappa=0.74$. It would be interesting to see if such dependences could be analytically foreseen, without any simulation and without any assumption on the tiling problem.

Let us consider first the simpler case of a single parameter $x$ ($D_{\mathbb{P}}=1$). One then assumes that the match is well approximated by its quadratic expansion around a given template located at $x=0$. Thus, the efficiency area ${\mathbb{E}}$ is bounded by the equation:

\begin{equation}
\nonumber
1 \, - \, k x^2 \, \ge \, MM
\end{equation}
with $MM$ the minimal match and $k$ the metric coefficient. ${\mathbb{E}}$ is a segment $\left[-x_{\text{max}};x_{\text{max}}\right]$ with $x_{\text{max}} = \sqrt{(1-MM)/k}$. Assuming a uniform distribution of physical signals in this area, it is easy to compute the mean fraction of recovered SNR:

\begin{equation}
\overline{\phi} \; = \; \frac{ 1 }{ 2 \, x_{\text{max}} } \; \int_{ -x_{\text{max}} }^{ x_{\text{max}} } \, dx \, ( 1 - k x^2 ) \; = \; 1 \, - \, \frac{ 1 \, - \, MM }{ 3 }
\end{equation}
With $MM=97\%$, one gets $\overline{\phi}=1\%$ which is very close to the value of $0.9\%$ achieved in our simulations. It is worth noting that the result is independant of $k$ (and so of the particular efficiency area considered), which validates the assumption of computing the mean value of the fraction of SNR recovered.

One can estimate $\overline{{\mathfrak{L}}}$ in the same way:

\begin{equation}
\overline{{\mathfrak{L}}} \; = \; 1 \, - \, \frac{ 1 }{ 2 \, x_{\text{max}} } \; \int_{ -x_{\text{max}} }^{ x_{\text{max}} } \, dx \, ( 1 - k x^2 )^3 \; = \; ( 1 \, - \, MM ) \; - \; \frac{ 3 ( 1 \, - \, MM )^2 }{ 5 } \; + \; \frac{ ( 1 \, - \, MM )^3 }{ 7 } 
\end{equation}
The choice of $MM=97\%$ gives $\overline{{\mathfrak{L}}}=2.9\%$, while with $MM=85\%$, one has $\overline{{\mathfrak{L}}}=13.7\%$. These values exceed the results of the numerical simulations (respectively 2.6\% and 10.7\%) but are quite comparable.

How do these results change with a two-dimensional parameter space $(x,y)$ -- which is indeed the situation considered in this article? By a proper choice of coordinates, one can -- like for the 1D-case -- assume that the chosen template is located at $(0,0)$ and that the efficiency area is defined by the equation of an ellipsoid written in its simplest form:

\begin{equation}
\nonumber
\frac{ x^2 }{ a^2 } \, + \, \frac{ y^2 }{ b^2 } \; \le \; 1 \, - \, MM
\end{equation}
Straightforward calculations allow one to compute both the mean fraction of recovered SNR $\overline{\phi}$ and the mean loss of events $\overline{{\mathfrak{L}}}$:

\begin{eqnarray}
\overline{\phi} &=& 1 \, - \, \frac{ 1 \, - \, MM }{ 2 } \\
\overline{{\mathfrak{L}}} &=& 1 \, - \, \frac{ 1 \, - \, MM^4 }{ 4 ( 1 \, - \, MM ) } \; = \; 1 \, - \, \frac{ 1 \, + \, MM \, + \, MM^2 \, + \, MM^3 }{ 4 }
\end{eqnarray} 
Like for the one-dimensional case, these expressions do not depend on the particular template and are thus estimators of the mean values of these quantities. With $MM=97\%$, one gets values higher than those computed numerically: $\overline{\phi}=1.5\%$ and $\overline{{\mathfrak{L}}}=4.4\%$ respectively. The discrepancy between these numbers and those given in the case $D_{\mathbb{P}}=1$ is higher, even if the tiling studied here is 2-dimensional.

\begin{figure}
\centerline{\epsfig{file=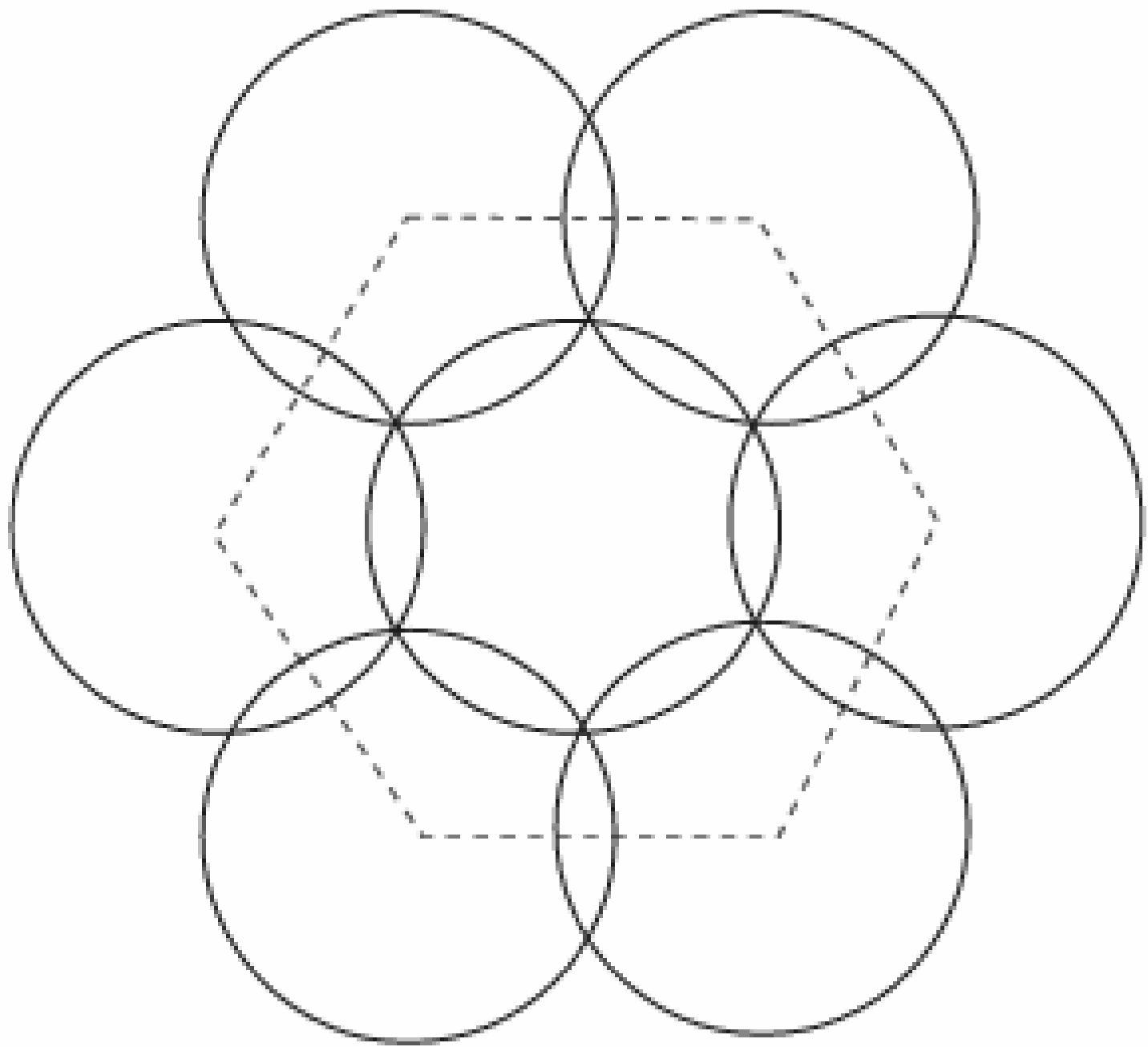,width=20cm}}
\caption{Optimal tiling of an infinite plane by identical circles: the centers belong to an hexagonal lattice. The overlapping is minimal: ~21\%}
\label{fig:optimal_circle_tiling}
\end{figure}

The effect of the overlap between close templates for $D_{\mathbb{P}}=2$ has been checked with the fully simplified model presented in Figure \ref{fig:optimal_circle_tiling}: efficiency areas are identical disks of a given radius. In this case, a more accurate expression of ${\overline{\mathfrak{L}}}$ can be computed analytically by selecting for each physical signal the closest template, i.e. the one allowing to recover the highest fraction of SNR. But the gain is very small: the fraction of false dismissals decreases only by about 1-2\% and remains in all cases higher than for the 1D-model. In fact, one could not expect much more from this refinement. Figure \ref{fig:optimal_circle_tiling} shows that only a fraction ($1-2\pi/3\sqrt{3}\approx 21\%$) of the physical signals are affected by this improvement and that these particular signals have the worst matches.

Finally, it appears that the good performances in detection efficiency of the template bank are the consequence of two aspects: firstly, the parameter space considered in this study is much more complex than the ideal case of a parameter space where one could use a uniform lattice of templates, and secondly the generated tilings remain redundant despite all the steps used to reduce it as much as possible. There is certainly still room for future improvements for the geometrical tiling algorithm.  

The comparison of the numerical data with the analytical parameterization computed for the model $D_{\mathbb{P}}=2$ (2D) is also shown on Figure \ref{fig:tiling_characteristics}. On this plot, it is clear that it overestimates the SNR losses with respect to the simulated tiling performances; yet, it gives values closer to real data than the $1-MM^3$ curve. Thus, the fractions of false dismissals are also higher.

%================================================================
\section{Conclusion}
%================================================================

This paper presents a method to tile any two dimensional parameter space $\mathbb{P}$ with a given maximal loss in SNR. The template list is build in two steps: the first one is an iterative algorithm which provides a complete coverage of $\mathbb{P}$ without any area left uncovered. As the computed set of filters can be redundant, a second step including two 'cleaning' steps allows one to significantly decrease the number of templates by dropping those which are useless. This algorithm has been used for the lattice of templates needed to look for damped sine signals with matched filtering. Even if the number of filters involved is by some orders of magnitude below those computed for the inspiralling binaries search, this choice of waveform allowed us to strongly test the capability of the algorithm as the ellipse characteristics show large variations in $\mathbb{P}$ or, equivalently that $\mathbb{P}$ is curved. Moreover, the algorithm using these templates can be directly implemented in an on-line filtering.

As it is not possible to choose the optimal tiling of a given parameter space, the final configuration of filters is computed by mixing various sets of templates computed with different initial conditions (the location of the initial filter used to develop the iterative algorithm) and by applying to the global list the reduction procedure. This kind of 'average' decreases the number of filters by about 15\% with respect to an unique procedure. The number of templates finally computed is comparable to the analytical estimations and Monte-Carlo simulations show that the set of filters fulfills the initial requirement: minimizing the loss in SNR in the whole parameter space. 

In the damped sine case, the study of the sets of templates shows that the effective loss of events $\overline{\mathfrak{L}}$ (as a function of $MM$) is much less than its usual estimation: $\overline{\mathfrak{L}}$ approximately behaves as $1 - MM^{0.74}$ instead of $1 - MM^3$. With the usual prescription of 10\% for the loss, the number of templates can be decreased by a factor close to 4. Seen in a reverse way, for a given $MM$ prescription, the SNR fraction recovery is on average much higher than $MM$: 99.1\% for $MM=97\%$ as an example. Even if the exponent value depends on the specific waveform details, such a behavior should be expected for the inspiralling binary case. This feature can allow us either to decrease the number of templates (and then the computing power) or perhaps obtain a more robust (with respect to noise for example) tiling of the parameter space.

Finally, simple parameterizations allowing one to predict the dependence in $MM$ of the mean fraction of recovered SNR $\overline{\phi}$ and of the false dismissal fraction $\overline{\mathfrak{L}}$ are presented. They are derived independently of the particular tiling method studied in this paper and should thus be generic. They do not give the power-law scaling of $\overline{\mathfrak{L}}$ inferred from the results of our numerical simulations and they both overestimate the losses due to the finite lattice of templates w.r.t. the tilings we generated. It would be interesting to compare this situation with different tilings generated with other algorithms.

\vskip 1.5cm
\noindent Acknowledgment: the authors wish to thank the referee for his fruitful comments.

%============================================================================
% For figures
%============================================================================

%=========================================================================
\baselineskip = 0.5\baselineskip  
%=========================================================================

%=========================================================================

%=========================================================================

\end{document}